\documentclass{eptcs} 

\usepackage{mathrsfs}

\usepackage{rotating}

\usepackage{graphicx}
\usepackage{amssymb}
\usepackage{amsmath}

\usepackage[strings]{underscore}
\usepackage{xspace}
\usepackage{url}

\newcommand{\restrict}{\upharpoonright} 

\newcommand{\commentout}[1]{}
\newtheorem{theorem}{Theorem}
\newtheorem{proposition}{Proposition}

\newtheorem{definition}{Definition}

\newcommand{\agts}{\mathit{Agts}}

\newcommand{\be}{\begin{enumerate}}  
\newcommand{\ee}{\end{enumerate}}  

\newcommand{\bi}{\begin{itemize}}  
\newcommand{\ei}{\end{itemize}}  

\newcommand{\Prop}{V}
\newcommand{\vars}{\mathit{vars}}

\newcommand{\rimp}{\Rightarrow}
\newcommand{\dimp}{\Leftrightarrow}

\newcommand{\assgt}{\mathit{assgt}}

\newcommand{\ks}{M}

\newcommand{\vs}{\mathcal{M}}
\newcommand{\Ovars}{O}

\newcommand{\relss}{{\cal S}}
\newcommand{\sys}{{\cal I}}

\begin{document} 
\title{Optimizing Epistemic Model Checking Using Conditional Independence (Extended Abstract)
\thanks{
Work supported by US Air Force, Asia Office of Aerospace Research and Development, 
grant AFOSR FA2386-15-1-4057. 
Thanks to Xiaowei Huang and Kaile Su for some preliminary discussions and investigations
on the topic of this paper. 
An extended version of this paper with proofs and  additional information is available at \url{https://arxiv.org/abs/1610.03935}.
} }

\author{Ron van der Meyden
\institute{UNSW Sydney, Australia}
\email{meyden@cse.unsw.edu.au}
}
 
 \def\titlerunning{Optimizing Epistemic Model Checking Using Conditional Independence} 
\def\authorrunning{Ron van der Meyden}

\maketitle

\begin{abstract} 
This paper shows that conditional independence reasoning can be 
applied to optimize epistemic model checking, in which one 
verifies that a model for a number of agents operating with 
imperfect information satisfies a formula expressed in a modal multi-agent logic of knowledge. 
The optimization has been implemented in the epistemic model checker MCK.  
The paper reports experimental results demonstrating that it can yield multiple orders of magnitude
performance improvements. 
\end{abstract}

\section{Introduction} 

\noindent  
Epistemic model checking \cite{mck} is a technique for the verification of 
information theoretic properties, stated in terms of a modal logic of knowledge, 
in systems in which multiple agents operate with imperfect information of
their environment.  It has been applied to settings that include 
diagnosis \cite{BozzanoCGT15}, and reasoning in game-like settings \cite{HuangMM11,HuangRT13,DitmarschHMR06}, 
concurrent hardware protocols \cite{BaukusM04} and security protocols \cite{BatainehMeyden11,BoureanuCL09,MS}. 

The contribution of the present paper is to demonstrate that  
conditional independence techniques from the Bayesian Net literature \cite{KollerFriedman,pearlprobbook,Darwiche97} 
can be applied in 
the context of epistemic model checking. We develop a generalization of  these techniques for 
a multi-agent modal logic of knowledge, that enables
model checking computations for this logic to be optimized by reducing the
number of variables  that need to be included in data structures
used by the computation. 

We have implemented the technique in the epistemic model checker MCK \cite{mck}. 
The technique developed can be applied for other semantics and algorithms, but 
we focus here on agents with \emph{synchronous perfect recall} and model check the 
reduced representation using binary decision diagram techniques. 
The synchronous perfect recall semantics presents
the most significant challenges to the computational cost of 
epistemic model checking, since it 
leads to a rapid blowup in the number of variables that need
to be handled by the symbolic model checking algorithms. 

The paper presents experimental results that demonstrate 
that the conditional independence optimization yields
very significant gains in the performance of 
epistemic model checking. Depending on the 
example, the optimization yields a speedup 
as large as four orders of magnitude. 
Indeed, it can yield linear growth rates in  computation time on examples
that otherwise display an exponential growth rate. 
It adds significantly to the scale of the examples
that can be analyzed in reasonable time, 
increasing both the number of agents that can be 
handled, the length of their protocols, and 
the size of messages they communicate.

\section{Background: Epistemic Logic} 
\label{sec:background}

We begin by recalling some basic definitions from epistemic logic and 
epistemic model checking. We use {\em epistemic variable structures}, 
a particular concrete representation of Kripke structures (it can be shown that there is 
no loss of generality). We show how these structures arise in a multi-agent setting in 
which each agent's behaviour is described by a program.

Let $\Prop$ be a set of atomic propositions, which we also call {\em variables}. 
An assignment for a set of variables $V$ is a mapping $\alpha: V\rightarrow \{0,1\}$. 
We write $\assgt(V)$ for the set of all assignments to variables $V$.
We denote the restriction of a function $f:S\rightarrow T$ to a subset $R$ of the domain $S$ 
by  $f\restrict R$. 

The syntax of epistemic logic for a set $\agts$ of agents is given by the grammar 
$$ \phi ::= p ~|~ \neg \phi ~|~ \phi \land \phi~|~ K_i\phi $$ 
where $p\in \Prop$ and $i\in \agts$.  That is, the language is 
a modal propositional logic with a set of modalities $K_i$, 
such that $K_i\phi$ means, intuitively, that the agent $i$ knows that $\phi$.   
We freely use common abbreviations from propositional logic, 
e.g., we write $\phi_1 \lor \phi_2$ for $\neg (\neg \phi_1 \land \neg \phi_2)$
and $\phi_1 \rimp \phi_2$ for $\neg \phi_1 \lor \phi_2$ and 
$\phi_1 \dimp \phi_2$ for $(\phi_1 \rimp \phi_2) \land (\phi_2 \rimp \phi_1)$. 
We write $\vars(\phi)$ for the set of variables occurring in the formula $\phi$.

Define an {\em epistemic variable structure} over a set of variables $V $ to be a 
tuple $\vs = (A, O,V)$ where   $A \subseteq \assgt(V)$ and 
$O = \{O_i\}_{i\in \agts}$ is a collection of sets of variables 
$O_i\subseteq V$, one for each agent $i$.
Intuitively, such a structure is an alternate representation of an epistemic Kripke structure, 
where the indistinguishability relation for an agent is specified by means of a set of variables
observable to the agent.  The elements of $A$ correspond to the worlds of this 
Kripke structure. 
The relation $\sim_i$ on worlds for agent $i$ is defined by 
$u \sim_i v$ when $ u \restrict O_i = v\restrict O_i$.

The semantics of epistemic logic is given by a ternary relation $\vs,w \models \phi$, 
where $\vs = (A, O,V)$ is an epistemic variable structure, 
 $w\in A$ is a world of $\vs$, 
and $\phi$ is a formula. The definition is given recursively, 
by 
\be
\item $\vs,w \models p$ if $w(p) = 1$, for $p \in \Prop$,  
\item $\vs,w \models \neg \phi$ if not $\vs,w \models \phi$,
\item $\vs,w \models \phi_1\land \phi_2$ if $\vs,w \models \phi_1$ and $\vs,w \models \phi_2$, 
\item $\vs,w \models K_i \phi$ if  $\vs,u \models \phi$ for all worlds $u\in A$ with $w\sim_i u$. 
\ee  
Intuitively, the clause for the operator $K_i$ says that $K_i\phi$ holds when $\phi$ is true at
all worlds that the agent considers to be possible.  
We write $\ks \models \phi$ when $\ks,w\models \phi$ for all worlds $w\in W$.

\newcommand{\atl}{\langle} 
\newcommand{\atr}{\rangle} 
\newcommand{\skp}{\mathit{skip}}

In the context of model checking, one is interested in analyzing a model 
represented as a program. We now show how programs generate a Kripke structure that serves as their semantics. 
We work with a very simple straightline programming language in which a multi-agent scenario is 
represented by each of the agents running a protocol in the context of an environment. 
The syntax and operational semantics  of this language  is shown in Figure~\ref{fig:lang}. 

Intuitively, all variables (represented by non-terminal $v$) in this fragment are boolean, and $e$ represents a boolean expression. 
Code $C$ consists of a sequence of assignments and randomization statements $rand(v)$, which assign a random value to $v$. 
(In a probabilistic interpretation, the random value would be drawn from a uniform 
distribution, but for our purposes in epistemic model checking, we
interpret this operation as nondeterministically selecting a value of either $0$ or $1$.)
Non-terminal $a$ 
represents an atomic action, either the skip statement $\skp$, or an atomic statement 
$\atl C\atr$  consisting of code $C$ that executes without interference
from code of other agents.  An agent protocol $P$ consists of a sequence 
of atomic actions: protocol $\epsilon$ represents termination, and 
is treated as equivalent to $\skp;\epsilon$ to capture that a terminated
agent does nothing while other agents are still running. 
A \emph{joint protocol} $J$, is 
represented by a statement of the form $P_1 ~||~ \ldots ~||~P_n~\Delta~C_E$, 
and consists of a number of agent protocols $P_1,\ldots,P_n$, 
running in the context of an environment represented by code $C_E$. 

There are two relations in the operational semantics. 
States $s$ are assignments of boolean variables to boolean values, 
and we write $e(s)$ for the value of boolean expression expression $e$ 
in state $s$. The binary relation $\rightarrow_0$ on configurations of type $(s,C)$
represents \emph{zero-time} state transitions, which do not change
the system clock. The binary relation $\rightarrow_1$ 
on configurations of type $(s,J)$  represents state transitions 
corresponding to a single clock tick. 
Thus, 
$C\rightarrow_0^* \epsilon$ represents that code $C$  runs to termination in time $0$. 
In a single tick transition represented by $\rightarrow_1$, we take the next 
atomic action $a_i = \atl C_i\atr$ from each of the agents, 
and compose the code $C_i$ in these actions  with the code from the environment 
$C_E$ to form the code $C= C_1;\ldots C_n; C_E$. The single step transition
is obtained as the result of running this code $C$ to termination in
zero-time. 

\begin{figure} 
$$\begin{array}{l} 
e ::= v ~|~ \neg v~| ~v\land v ~|~ v \lor v~| ~ \ldots \\ 
C::= \epsilon ~|~ v := e ; C~|~ rand(v) ; C \\ 
a ::= \atl C\atr ~|~ \skp \\ 
P ::= \epsilon ~|~   a ; P\\ 
J ::= P~ ||~ ... ~||~ P~\Delta ~ C 
\end{array} 
$$

$$\begin{array}{ccc}
 (s, v:=e;C) \rightarrow_0 (s[e(s)/v], C) & \quad \quad & 
(s, \skp;C) \rightarrow_0 (s,C) \\[10pt]
(s, rand(v);C) \rightarrow_0 (s[0/v], C) & & 
(s, rand(v);C) \rightarrow_0 (s[1/v] ,C) 
\end{array} 
$$

$$\begin{array}{c} 
a_1 = \atl C_1 \atr  ~ \ldots  ~a_n = \atl C_n \atr ~~~~C= C_1 ; \ldots ; C_n; C_E~~~~(s, C) \rightarrow_0^* (t, \epsilon)\\  
\hline 
(s,  ~a_1 ;P_1 ~||~ \ldots  ~||~ a_n;P_n ~ \Delta~ C_E) \rightarrow_1  (t, ~P_1 ~|| ~\ldots ~||~ P_n  ~\Delta~ C_E )
\end{array} 
$$
\caption{Syntax and Operational Semantics of Programs\label{fig:lang}} 
\end{figure} 

A {\em system}  is represented using this programming language by means of a
tuple $\sys = (J,I,Q)$, where $J$ is a joint protocol for $n$ agents, 
$I$ is a boolean formula expressing the initial condition, 
and $Q$ is a tuple of $n$ sets of variables, with $Q_i$ representing
the variables observable to agent $i$. 

Given a maximum running time $n$, a system $\sys=(J,I,Q)$ is associated to an epistemic variable structure 
$\vs_n(\sys)=\langle A,O, V\rangle$ as follows. A 
\emph{run} of length $n$ of the system is a sequence of states $r= s_0, s_2, \ldots, s_n$, where 
$s_0$ satisfies the initial condition $I$ and 
$(s_0,J) \rightarrow_1 (s_1,J_1) \rightarrow_1 \ldots \rightarrow_1 (s_n,J_n) $ for some $J_1, \ldots, J_n$.  
If $U$ is the set of variables appearing in $J$, 
we define $V$ to be the set of \emph{timed variables}, 
i.e., the set of variables $v^t$ where $0\leq t \leq n$.  
We take  $A$  to be the set of assignments $\alpha_r$ to variables $V$ 
derived from runs $r$ by $\alpha_r(v^t) = s_t(v)$ when $v\in U$ and $0\leq t \leq n$. 
For the perfect recall semantics, which is our focus in this paper, 
we define the observable variables $O_i$ for agent $i$ 
to be the set of timed variables $v^t$ where $v\in Q_i$ and $0 \leq t\leq n$.

\section{Example: Dining Cryptographers} 
\label{sec:exampledc} 
We illustrate epistemic model checking and the optimizations developed in this paper 
using Chaum's Dining Cryptographers Protocol \cite{chaum},  a security protocol whose aim is 
to achieve an anonymous broadcast. This protocol, both in its basic form, as well as an extension that is
more generally applicable, has previously been analysed
using epistemic model checking \cite{MS,AlBatainehMeyden10}. Chaum introduces the protocol  with the following story: 

\begin{quote}
Three cryptographers are sitting down to dinner at their favourite restaurant.
Their waiter informs them that arrangements have been made with the maitre d'hotel
for the bill to be paid anonymously. One of the cryptographers might be paying for
the dinner, or it might have been NSA (U.S. National Security Agency). The three
cryptographers respect each other's right to make an anonymous payment, but they
wonder if NSA is paying. They resolve their uncertainty fairly by carrying out the following protocol:

Each cryptographer flips an unbiased coin behind his menu, between him and the cryptographer on his right, so that only the two of them can see the outcome. Each cryptographer then states aloud whether the two coins he can see--the one he flipped and the one his left-hand neighbor flipped--fell on the same side or on different sides. If one of the cryptographers is the payer, he states the opposite of what he sees. An odd number of differences uttered at the table indicates that a cryptographer is paying; an even number indicates that NSA is paying (assuming that the dinner was paid for only once). Yet if a cryptographer is paying, neither of the other two learns anything from the utterances about which cryptographer it is.
\end{quote}

\newcommand{\paid}{\mathit{paid}} 
\newcommand{\coin}{\mathit{coin}}
\newcommand{\leftcoin}{\mathit{left}}
\newcommand{\say}{\mathit{say}}
\newcommand{\pinit}{\mathit{pinit}}
  
The solution generalizes to any number $n$ of cryptographers $C_0, \ldots , C_{n-1}$ 
at the table.  We may represent the protocol by means of the following program for 
cryptographer $i$, who is assumed to have a boolean variable $\paid_i$ 
that indicates whether (s)he is the payer. (The program starts running from an initial state
in which the  constraint $\bigvee_{0\leq i < j \leq  n-1} \neg( \paid_i\land \paid_j)$  is satisfied.) 
We write $\oplus$ for the exclusive-or. 

\begin{tabbing} 
nnn \= \kill 
$C_i$: \\
Observed variables: $\paid_i$, $\coin_i$, $\leftcoin_i$, $\say_0, \ldots , \say_{n-1}$\\ 
Protocol: \\ 
\> $rand(\coin_i)$ ; \\ 
\> $\leftcoin_{i+1~mod~n} := \coin_i$ ; \\
\> $\say_i := \paid_i \oplus \coin_i \oplus \leftcoin_i$ 
\end{tabbing} 
All variables take boolean values. 
Each cryptographer is associated with a set of variables, whose values they are  
able to observe at each moment of time. Note that a cryptographer may write to 
a variable that they are not able to observe. In particular, $C_i$ writes to the variable 
$\leftcoin_{i+1~mod~n}$ that is observed only by $C_{i+1~mod~n}$.

\begin{figure} 
\centerline{\includegraphics[height=10cm]{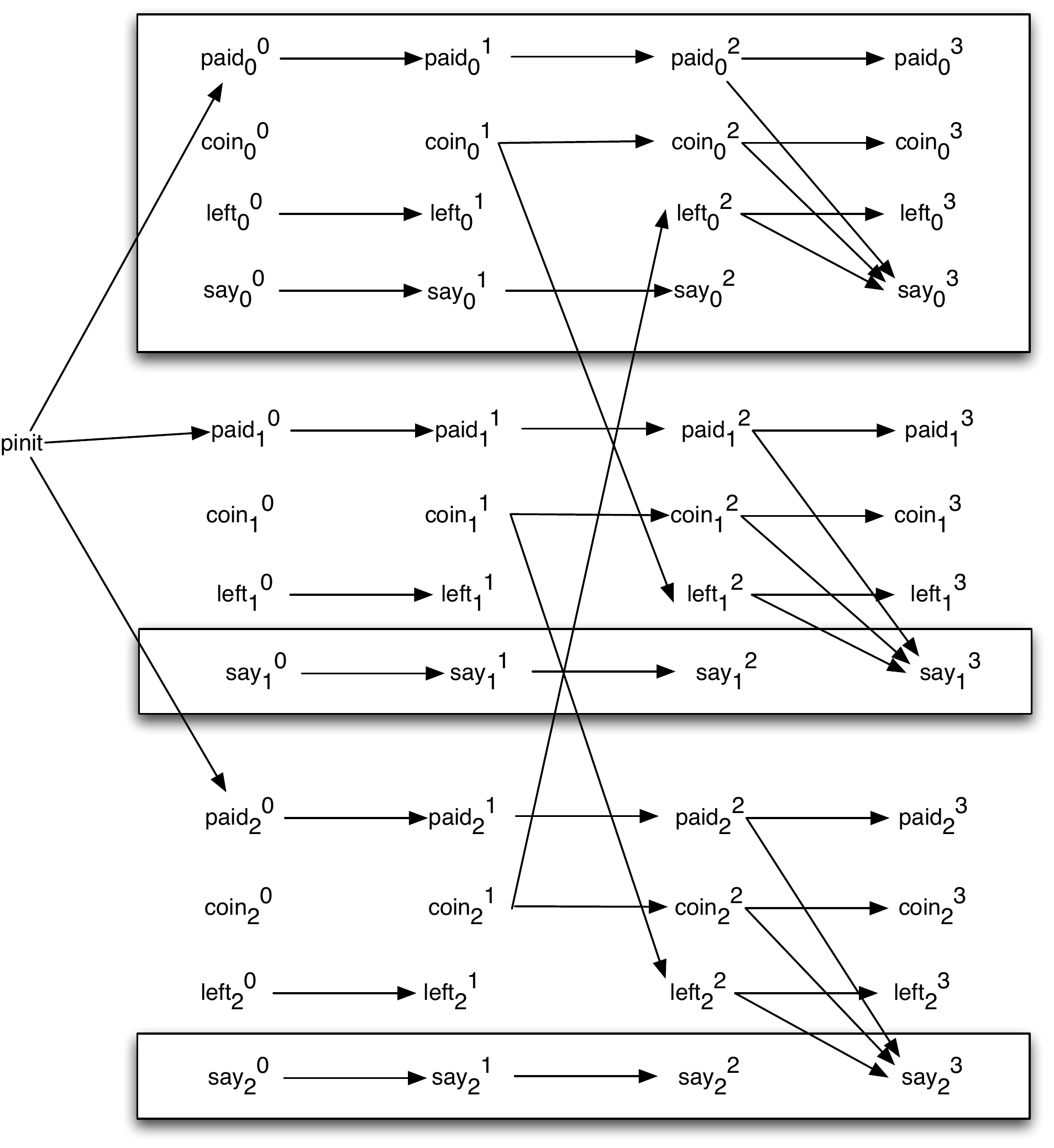}}
\caption{Timed-variable dependency graph after program unfolding \label{fig:initgraph}} 
\end{figure}

We will work with {\em dependency networks} that  show how the values of 
variables change over time. 
The  DC protocol runs for 4 ticks of the clock, 
(time 0 plus one tick for each step in the protocol), so we have instances $v^0 \ldots v^3$ 
of each variable $v$. Figure~\ref{fig:initgraph} shows the dependencies between 
these instances. The figure is to be understood as follows: a variable $v^t$ takes a value
that directly depends on the values of the variables $u^{t-1}_1 \ldots u^{t-1}_n$ such that there is 
an edge from $u^{t-1}_j$ to $v^t$.  Additionally, there is a dependency between the initial 
values $\paid_i^0$ captured using a special variable $p_{init}$. (We give a more formal presentation of such dependency structures  below.)  
The observable variables for agent $C_0$ have been indicated by rectangles: timed variables inside
these rectangles are observable to $C_0$. 

\section{Valuation Algebra} \label{sec:valalg} 

Shenoy and Shafer \cite{Shenoy89,ShenoyShafer90}  have developed a general axiomatic formalism that captures the 
key properties that underpin the correctness of optimization methods used for a variety of 
uncertainty formalisms. In particular, it has been shown that this 
formalism allows for a general explanation of variable elimination 
algorithms and the notion of conditional independence used in 
the Bayesian Network literature \cite{KollerFriedman}, and applies
also in other contexts such as Spohn's theory of \emph{ordinal conditional 
functions} \cite{Spohn88}.  There is a close connection also to ideas 
in database query optimization \cite{Maier83} and operations research \cite{BB72}. 
We show here that Shenoy and Shafer's general axiomatic framework 
applies to 
epistemic model checking. 
This will enable us to apply the variable elimination algorithm  
to derive techniques for optimizing epistemic model checking. 

\newcommand{\Vars}{\mathit{Vars}} 
\newcommand{\Valns}{\Phi} 
\newcommand{\dom}[1]{\mathit{dom}(#1)} 
\newcommand{\val}[1]{\Omega_#1}

\newcommand{\comb}{\otimes} 
\newcommand{\marg}{\downarrow} 
\newcommand{\Doms}{D} 
\newcommand{\valo}{s}
\newcommand{\valt}{t}
\newcommand{\idty}{e}


We begin by presenting Shenoy and Shafer's framework, following \cite{KohlasShenoy}. 
Let $\Vars$ be a set of variables, with each $v \in \Vars$ taking values in a set $\val v$. 
For a  set $X$ of variables, the set  $\val{X} = \Pi_{x\in X} \val{x}$
is called the \emph{frame} of $X$. Elements of $\val{X}$  are called 
\emph{configurations} of $X$. In case $X=\emptyset$, the set 
$\val{X}$ is interpreted as $\{\langle\rangle\}$, i.e., the set 
containing just the empty tuple. We write $\Doms$ for ${\cal P}(\Vars)$.

A \emph{valuation algebra} is a tuple $\langle \Phi, \mathit{dom}, e, \comb,\marg\rangle$, 
with components as follows.  A state of information is represented in valuation algebra 
by a primitive notion called a \emph{valuation}. Component $\Phi$ is a set, the set of all valuations, 
and $\mathit{dom}$ is function from $\Phi$ to $\Doms$. 
Intuitively, for each valuation $s\in \Phi$, the domain $\dom{s}$ is the set of 
variables that the information is about. For a set of variables $X$, we write $\Valns_X$
for the set of valuations  $s$ with $\dom{s} =X$.  
Component $\idty$ gives an element $\idty_X\in \Phi_X$ for each $X\in D$. 
A valuation algebra also has two operations $\comb: \Valns \times \Valns \rightarrow \Valns$ (combination) and $\marg: \Valns\times \Doms \rightarrow \Valns$
(marginalization), 
with  $\comb$, intuitively, representing the combination of two pieces of 
information, and $\marg$   used to restrict a piece of information to a given set
of variables.  Both are written as infix operators. From marginalization, another operator $-: \Valns\times \Vars \rightarrow \Valns$ 
called {\em variable elimination} can be defined, by $\valo^{-x} = \valo \marg (\dom{\valo} \setminus \{x\}) $. 

These operations are required to satisfy the following conditions: 
\be
\item[VA1.]  {\em Semigroup.} $\comb$ is associative and commutative. 
For all $X\in  \Doms$ and  all
$\valo\in \Valns_X$, we have $\valo\comb \idty_X = \idty_X\comb \valo = \valo$. 

\item[VA2.] {\em Domain of combination.} For all $\valo,\valt\in\Valns$, $\dom{\valo \comb\valt} = \dom{\valo} \cup \dom{\valt}$. 

\item[VA3.]{\em Marginalization.} For $\valo \in \Valns$ and $X,Y\in \Doms$, the following hold:  \\
$
\valo\marg X = \valo \marg X\cap \dom{\valo} \quad\quad 
\dom{\valo \marg X} = X\cap \dom{\valo} \quad\quad 
\valo \marg \dom{\valo} = \valo ~.
$ 

\item[VA4.] {\em Transitivity of marginalization.} For $\valo \in \Valns $, and $X\subseteq Y\subseteq \Vars$, \\
$ (\valo \marg Y)\marg X = \valo\marg X ~.$

\item[VA5.]  {\em Distributivity of marginalization over combination.} 
For $\valo,\valt\in \Valns$, with $ \dom{\valo} = X$, we have 
$ (\valo \comb \valt) \marg X = \valo \comb (\valt\marg X)~.$

\item[VA6.] {\em Neutrality.} For $X, Y\in \Doms$, 
$ \idty_X \comb \idty_Y = \idty_{X\cup Y}~. $ 

\ee 

\newcommand{\Fus}{\mathit{Fus}}
\newcommand{\bigcomb}{\bigotimes}

A key result that follows from these axioms, called the 
\emph{Fusion Algorithm} \cite{Shenoy92}, exploits 
Distributivity of Elimination over Combination to give 
a way of computing the result of a marginalization operation 
applied to a sequence of combinations, by 
\emph{pushing in} variable eliminations over elements of the 
combination that do not contain the variable. 

For a finite set $S = \{\valo_1, \valo_2, \ldots, \valo_k\} \subseteq \Valns$, 
write $\comb S$ for $\valo_1 \comb \valo_2 \comb \ldots \comb \valo_k$. 
We define the \emph{fusion} of $S$ via $x\in \Vars$ to be 
the set $\Fus_x(S) = \{ (\comb S_+)^{-x}\} \cup S_-~$
where we have partitioned $S$ as $S_+\cup S_-$, such that 
$S_+$ is the set of $\valo\in S$ with $x\in \dom{\valo}$, 
and $S_-$ is the set of $\valo\in S$ with $x\not \in \dom{\valo}$. 
That is, in the fusion of the set $S$ with respect to $x$, we combine all the
valuations with $x$ in their domain, and then eliminate $x$, and
preserve all valuations with $x$ not in their domain. 

Suppose we are interested in computing 
$(\comb S)\marg X$, for $S$ a finite set of valuations, 
and $X\subseteq \Vars$. The Fusion Algorithm 
achieves this by repeatedly applying the fusion 
operation, using some ordering of the variables in $X$. 
We write $\dom{S}$ for $\dom{\comb S} = \bigcup \{ \dom{\valo}~|~ \valo \in S\}$

\begin{theorem}[\cite{Shenoy92}] 
Let $S$ be a finite set of valuations, and $X\subseteq \Vars$. 
Suppose  $\dom{S} \setminus X = \{x_1,x_2, \ldots , x_n\}$. 
Then 
$ (\comb S) \marg X = \comb \Fus_{x_n}( \ldots (\Fus_{x_1}(S))) ~.$ 
\end{theorem} 

Each ordering of the variables $x_1 \ldots x_n$ gives 
a different way to compute $(\comb S) \marg X$. 
A well chosen order can yield a significant optimization 
of the computation, by keeping the domains of the intermediate valuations in the sequence of fusions small. 
Finding an optimal order may be  computationally complex, but there exist heuristics that 
produce good orders in practice \cite{Olmsted,Kong}.

We now show that the relational structures that underly Kripke structures are associated  with algebraic 
operations that satisfy the conditions VA1-VA6. It will follow from this 
that the Fusion algorithm can be applied to these structures. 

\newcommand{\AllProp}{{\cal V}}
\newcommand{\rels}{s}

Let $\AllProp$ be the set of all variables. 
Values in the algebra will be \emph{relational structures} of the form $\rels = (A,V)$, 
where $V\subseteq \AllProp$ and $A\subseteq \assgt(V)$. 
The domain of a relational structure is defined to be its set of variables, 
i.e. if $\rels = (A,V)$ then $\dom{\rels} = V$. 
We define the identities $\idty_X$ and operations $\comb$  of combination and $\marg$ of marginalization as follows. 
Let $\rels_1 = (A_1, V_1)$ and $\rels_2 = (A_2, V_2)$ and $X \subseteq \AllProp$. 
Then 
\begin{itemize} 

\item $\idty_X = (\assgt(X),X)$,

\item $\rels_1 \comb \rels_2 = (A,V)$
where $V= V_1\cup V_2$,
and $A\subseteq \assgt(V)$ is defined by $\alpha \in A$ iff $\alpha\restrict V_1 \in A_1$ and $\alpha\restrict V_2 \in A_2$. 
\item $\rels_1 \marg X = (A,V)$ where 
$V = V_1 \cap X$, 
and $A = \{ \alpha\restrict X~|~ \alpha\in A_1\}$.
\end{itemize} 
To use terminology from relational databases, $\rels_1\comb \rels_2 $ is the join of
relations and $\rels \marg X $ is the projection of the relation $\rels$ onto attributes $X$.  
The following result is straightforward; these properties are well-known for 
relational algebra. 

\begin{proposition} 
The algebra of relational structures satisfies axioms VA1-VA6. 
\end{proposition}   

We may extend the operation of marginalization in this valuation algebra
to epistemic variable structures as follows. If $\vs= (A,O,V)$
is an epistemic variable structure and $X\subseteq V$, we define $\vs\marg X = (A',O',V')$
where $A' = \{\alpha\restrict X~|~ \alpha \in A\}$
and $O'_i = O_i \cap X$ for all $i\in \agts$ and $V' = V\cap X$. 
In general, this operation results in agents losing information, since their
knowledge is based on the observation of fewer variables. Below, we
identify conditions where knowledge is preserved by this operation.

\section{Conditional Independence and Directed Graphs}
 \label{sec:condind-dg}


Let $X,Y,Z\subseteq \Prop$ be sets of variables. The notion of conditional 
independence expresses a generalized type of independency relation. 
Variables $X$ are said to be conditionally independent of $Y$, given $Z$, 
if, intuitively, once the values of $Z$ are known, the values of $Y$ are 
unrelated to the values of $X$, so that neither $X$ not $Y$ gives
any information about the other. This intuition can be formalized for 
both probabilistic and discrete models. The following definition gives 
a discrete interpretation, related to the notion of 
\emph{embedded multivalued dependencies} from database theory \cite{Fagin77}.

\begin{definition} 
Let $A\subseteq \assgt(V)$ be a set of assignments  over variables $\Prop$ and 
let $X,Y,Z \subseteq \Prop$. We say that $A$ satisfies the conditional independency
$X \bot Y|Z$, and write $A \models X\bot Y|Z$, if 
for every pair of worlds $u,v \in A$ with $u\restrict Z = v\restrict Z$, there exists 
$w \in A$ with $w\restrict X\cup Z = u \restrict X\cup Z$ and $w\restrict Y\cup Z  = v\restrict Y\cup Z$. 
For an epistemic variable structure $\vs= (A,O,V)$, we write $\vs  \models X\bot Y|Z$ if 
$A \models X\bot Y|Z$. 
\end{definition} 

Conditional independencies can be deduced from graphical representations of models. 
Such representations have been used in the literature on Bayesian 
Nets \cite{pearlprobbook,KollerFriedman}, and have also been applied in propositional reasoning \cite{Darwiche97,Darwiche98}. 
The following presentation is similar to \cite{Darwiche97} except that we work with relations over arbitrary domains rather than 
propositional formulas. 

\newcommand{\arr}{\rightarrow}
\newcommand{\und}{-}
\newcommand{\parents}[1]{\mathit{pa}(#1)}
\newcommand{\ancestors}[1]{\mathit{An}(#1)}


The notion of d-separation \cite{pearlprobbook} provides a way to derive a set of 
independency statements from a directed graph $G$. We present here an equivalent formulation
from \cite{LDLL90}, that uses the notion of the \emph{moralized} graph $G^m$ of a directed graph $G$. 
The graph $G^m$ is defined to be the undirected graph obtained from $G$ by first adding an edge 
$u \und v$ for each pair $u,v$ of vertices that have a common child (i.e. such that 
there exists $w$ with $u\arr w$ and $v \arr w$), and then replacing all directed edges with undirected
edges. 
The set of \emph{parents} of a node $u$ is defined to be the set $\parents v = \{u\in V~|~ u\arr v\}$.  
 For a set of vertices $X$ of the directed graph $G$, we write $\ancestors{X}$ for the set of 
all vertices $v$ that are ancestors of some vertex $x$ in $X$ (i.e., such that there 
exists a directed path from $v$ to $x$). 
For a subset $X$ of the set of vertices of graph $G=(V,E)$, we define the restriction of
$G$ to $X$ to be the graph $G_X = (V\cap X,\{(u,v) \in E~|~ u,v\in X\})$.  
For disjoint sets $X,Y,Z$, we then 
have that \emph{$X$ is d-separated from $Y$ by $Z$} if all paths from $X$ 
to $Y$ in $(G_{\ancestors{X\cup Y\cup Z}})^m$ include a vertex in $Z$.

A \emph{structured model} for a valuation algebra $\langle \Phi, \mathit{dom}, e, \comb,\marg\rangle$
over variables $\Vars$,
is a tuple $M= \langle V,E, \relss \rangle$   
where $V\subseteq \Vars$ is a set of variables, component $E$ is a binary relation on $V$ such that $G_M = (V,E)$ is a dag,  and $\relss= \{\rels_v\}_{v\in V}$
 is a collection of values in $\Phi$ such that for each variable $v\in V$, we have 
 \begin{itemize} 
 \item 
 $\dom{\rels_v} = \{v\} \cup \parents v$, i.e. the domain of $\rels_v$  
 consists of $v$ and its parents in the dag, 
\item 
$\rels_v \marg {\parents v }= \idty_{\parents{v}}$.  
\end{itemize} 
Intuitively, the second constraint says that the relation $\rels_v$ does not constrain the parents of $v$: for each 
assignment of values to the parents of $v$, there is at least one value
of $v$ that is consistent. 

The following is a consequence of results in \cite{pearlprobbook,LDLL90,VP88}. 

\begin{proposition}
Suppose that $M= \langle V,E, \relss\rangle$   is a structured 
model 
and $X,Y,Z$ are disjoint subsets of the vertices $V$ of the directed graph $G = (V,E)$. 
If $X$ is d-separated from $Y$ by $Z$, then $\comb \relss \models X\bot Y|Z$. 
\end{proposition}

Structured models have an additional property that provides an 
optimization when eliminating variables: if a leaf node is
one of the variables eliminated from the combination of the nodes of the graph, 
then it can be  removed from the model without changing the result. 
This is captured in the following result. 

\begin{proposition} 
Suppose that $M= \langle V,E, \relss\rangle$   is a structured  model, 
let $X\subseteq V$ and let $v \in V\setminus X$ be a leaf node. 
Then $\comb \relss \marg X = \comb (\relss \setminus \{s_v\})\marg X$.  
\end{proposition} 

To apply these results for structured models to model checking epistemic logic, 
we use the following definition. We say that a structured model  $M= \langle V,E, \relss\rangle$ 
\emph{represents} the worlds of an epistemic variable structure $\vs=(A,O,U)$ 
if $V=U$ and $A = \comb \relss$. That is, the structured model 
captures the set of assignments making up the epistemic 
variable structure.

Consider the following formulation of the model checking problem: for an epistemic formula $\phi$, we wish to 
verify  $\vs\models \phi$ where $\vs = (A,\Ovars,V)$ is an 
epistemic variable  structure  with observable variables $\Ovars$, with worlds 
represented by a  structured model $M= \langle V,E, \relss\rangle$. 

A first idea for how to optimize this verification problem is to reduce the structure 
$\vs$ to the set of variables
$\vars(\phi)$, together with the sets $\Ovars_i$ for any operator $K_i$ in $\phi$. 
In fact, using the notion of  conditional dependence, 
it is often possible to identify a smaller set of  variables 
that suffices to verify the formula. The intuition for this 
is that some of the observed variables in $\Ovars_i$ may be independent of the 
variables in the formula, and moreover, information may be redundantly 
encoded in the observable variables.
The following definitions strengthen the idea of restricting to 
$\vars(\phi) \cup \Ovars$ by exploiting a sufficient condition for
the removal of observable variables.

\newcommand{\keep}{\kappa}

Say that $\keep$ is a \emph{relevance} function for a formula $\phi$ with respect to an
epistemic variable structure $\vs= (A, O, V)$ if it maps subformulas of $\phi$ to
subsets of the set of variables $V$, and satisfies the following  conditions: 
\be 
\item $\keep(p) = \{p\}$ for $p \in \Prop$, 
\item $\keep(\phi_1 \land  \phi_2 )=  \keep(\phi_1) \cup \keep(\phi_2)$, 
\item $\keep(\neg \phi_1) = \keep(\phi_1)$,  and 
\item $\keep(K_i\phi_1) = U_i \cup \keep(\phi_1)$, for some 
$U_i \subseteq O_i$ with $\keep(\phi_1) \cap O_i \subseteq U_i$ and 
$\vs \models (\keep(\phi_1) \setminus U_i )\bot (O_i \setminus U_i ) | U_i$. 
\ee  

In the final condition, $U_i$ can be any set. We note that a set $U_i$ satisfying the condition
can always be found. For, if we take $U_i = O_i$, then the condition states that 
 $\keep(\phi_1) \cap O_i \subseteq O_i$ and 
$M \models (\keep(\phi_1) \setminus O_i )\bot \emptyset | O_i$. 
Both parts of this statement are trivially true. In practice, we will want to 
choose $U_i$ to be as small as possible, since this will lead to stronger optimizations.%
\footnote{Since $(A\setminus C)\bot (B\setminus C)|C$ is equivalent to  
$A\bot B|C$, the independence condition could be more simply stated as 
$\keep(\phi_1) \bot O_i| U_i$. We work with the more complicated version because 
the algorithm for d-separation assumes disjoint sets.}

Note that $\phi$ is a subformula of itself, so in the domain of $\keep$. 
The following result says that satisfaction of $\phi$ is preserved when we marginalize 
to a superset of $\keep(\phi)$ for a relevance function $\keep$. 

\begin{theorem} 
Suppose that $\keep$ is a relevance function for $\phi$ with respect to epistemic variable structure $\vs$ 
and that $X$ is a set of variables with $\keep(\phi) \subseteq X \subseteq \dom{\vs}$. 
Then for all worlds $w$ of $\vs$, we have $\vs, w\models \phi$ iff $\vs\marg X, w \restrict X \models \phi$. 
\end{theorem}

{\bf Computing $\keep(\phi)$:}  The definition of $\keep$ provides a recursive
definition by which $\keep (\phi)$ can be calculated, with the exception that the 
case $\keep(K_i(\phi)) = U_i \cup \keep(\phi)$ allows for a choice of the set $U_i$, subject to 
the conditions $\keep(\phi) \cap O_i \subseteq U_i$ and 
$\vs \models (\keep(\phi) \setminus U_i )\bot (O_i \setminus U_i ) | U_i$. 
When the worlds of $\vs$ are represented by a structured relational model $M$,  
we show how to construct the \emph{minimal} set $U_i$ satisfying the stronger conditions
that  $\keep(\phi) \cap O_i \subseteq U_i$ and  $U_i$ d-separates  $\keep(\phi) \setminus U_i $ from  $O_i \setminus U_i $
in the directed graph $G$ associated with $M$.

Note $(\keep(\phi) \setminus U_i) \cup  (O_i \setminus U_i) \cup U_i= \keep(\phi) \cup O_i$
for any set $U_i$. Thus, the d-separation properties we
are interested in are computed in the moralized graph $H = (G_{\ancestors{O_i\cup \keep(\phi)}})^m$, 
which is independent of $U_i$. 
Let $U$ be the set of vertices $v\in O_i$ such that there exists 
a path in $H$
from a vertex $u \in \keep(\phi)\setminus O_i$ to $v$, 
with $v$ the first vertex on that path that is in $O_i$. 
The set $U$ can be constructed in linear time by a depth first search from $\keep(\phi)\setminus O_i$. 
Take $W = U \cup (\keep(\phi) \cap O_i)$.

\begin{proposition} 
$W$ is the smallest set satisfying the strengthened conditions for~$U_i$. 
\end{proposition}

Unfolding a program into a structured model tends to create a large number of 
timed variable instances whose associated value represents an equality between
two variables. Such instances can be eliminated by a simple transformation of the 
structured model.

For an assignment $\alpha$ with domain $V$, define $\alpha[y/x]$ to be the 
assignment $\alpha'$ with domain $(V\setminus\{x\})\cup \{y\}$ with 
$\alpha(y) = \alpha'(x)$ and $\alpha\restrict (\dom{s} \setminus \{x\} )= \alpha'\restrict (\dom{s} \setminus \{x\})$.
For a relational value $s$ and variables $x,y$ with $x \in \dom{s}$ and 
$y \not \in \dom{s}$, define $s[y/x]$ to be the relational value $t$ 
with $\dom{t} = (\dom{s} \setminus \{x\})\cup \{y\}$, 
consisting of all assignments $\alpha[y/x]$ for $\alpha \in s$. 
Intuitively, this is simply the relation $s$ with variable $x$ renamed to $y$. 

We extend this definition to structured relational models $M = (V,E,\relss)$ with $x,y\in V$, 
by defining $M[y/x] = (V', E', \relss')$ with 
$V' = V\setminus \{x\}$, and $E'=  E\cap (V' \times V')$, and 
$\relss = \{s'_v~|~v\in V'\}$, where $s'_v = s_v[y/x]$. 
In the following result, we write $\delta_{x,y}$ for the set of 
assignments $\alpha$ with domain $\{x,y\}$ and $\alpha(x) = \alpha(y)$. 

 \begin{proposition} 
Suppose that $M = (V,E,\relss)$ is a structured relational model with $x,y\in V$, and $\val{x} = \val{y}$, and 
$\parents{y} = \{x\} $ and $s_y = \delta_{x,y}$. Let $M[y/x] = (V', E', \relss')$. 
Then  $\comb \relss' = (\comb \relss) \marg V'$. 
 \end{proposition}

The definition furthermore extends to epistemic models 
$\vs = (A,O,V)$ with worlds represented by a 
structured relational model $M = (V, E, \relss)$. Let $M[y/x] = (V', E', \relss')$.  We define $\vs[x/y] = (A',O',V')$ where
$O' = \{O'_i\}_{i \in \agts}$ where $O'_i = O_i \cup \{y~|~x\in O_i\}$ for each $i \in \agts$, 
and $A' = \comb \relss'$. Note that $O'_i$ additionally makes variable $y$ visible to agent $i$ if 
$x$ was visible to $i$, in case this variable was not originally visible. 
 
 \begin{proposition} 
If $\val{x} = \val{y}$, and  $\parents{y} = \{x\} $ and $s_y = \{(x:a,y:a)~|~ a \in \val{x}\}$
then $\vs, \alpha \models \phi$ iff $\vs[y/x], \alpha[y/x] \models \phi[y/x]$. 
 \end{proposition}

The overall  optimized procedure for model checking  that we obtain from the above results 
uses the following steps: 
\be
\item We first unfold a program representation of the model into a structured relational model with symbolically represented values
and transform the query into a form that uses the timed instances variables in place of the 
original variables. This can be done in a way that builds in the equality optimization. 

\item We compute $\keep(\phi)$ using the algorithm above. 
\item We compute a symbolic representation of $\vs\restrict \keep(\phi)$, 
using the leaf node elimination optimization. 
\item We compute $\vs\restrict \keep(\phi) \models \phi$ in this representation
using a symbolic model checking algorithm.
\ee

\section{Example} \label{sec:dcopt} 

\begin{figure} 
\centerline{
\includegraphics[height=3in]{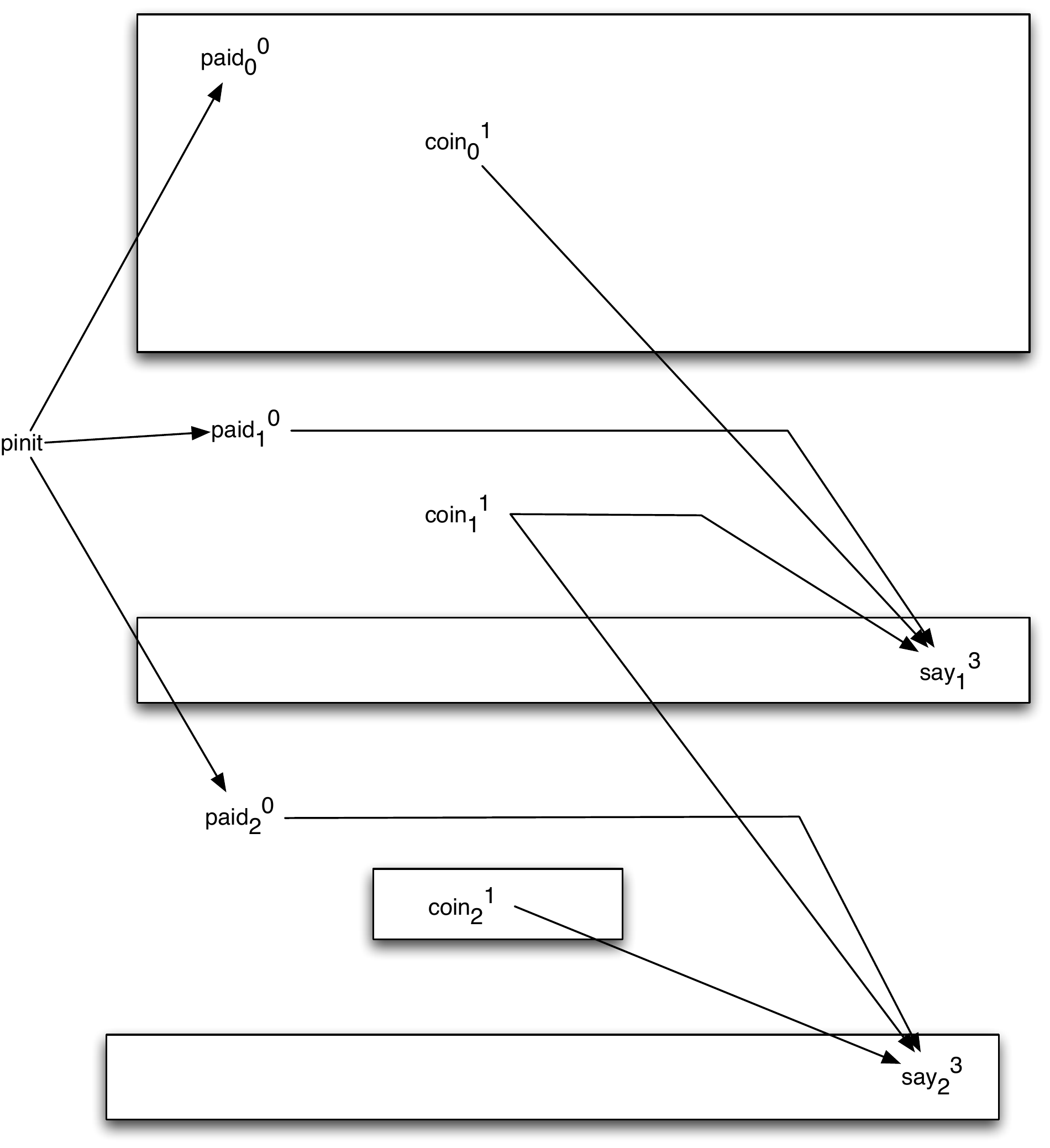}
}
\caption{Dining Cryptographers dependency graph after optimization\label{fig:optgraph}} 
\end{figure}

In the present section, we illustrate this procedure on the 
Dining cryptographers protocol. 
We consider the formula 
$$\phi = ( \neg \paid_0 \rimp K_0(\neg \paid_1 \land \neg \paid_2) \lor (K_0 (\paid_1 \lor \paid_1) \land  \neg K_0 \paid_1 \land \neg K_0 \paid_2)$$
evaluated at time 3. 
The dependency graph for the protocol was given above in Figure~\ref{fig:initgraph}. 
Figure~\ref{fig:optgraph} indicates the dependency graph that remains after we have 
applied the optimization procedure.   
The resulting formula is 
$$\phi_2 = ( \neg \paid^0_0 \rimp K_{0}(\neg \paid^0_1 \land \neg \paid^0_2) \lor (K_0 (\paid^0_1 \lor \paid^0_1) \land  \neg K_0 \paid^0_1 \land \neg K_0 \paid^0_2)$$

\begin{figure} 
\centerline{
\includegraphics[height=10cm]{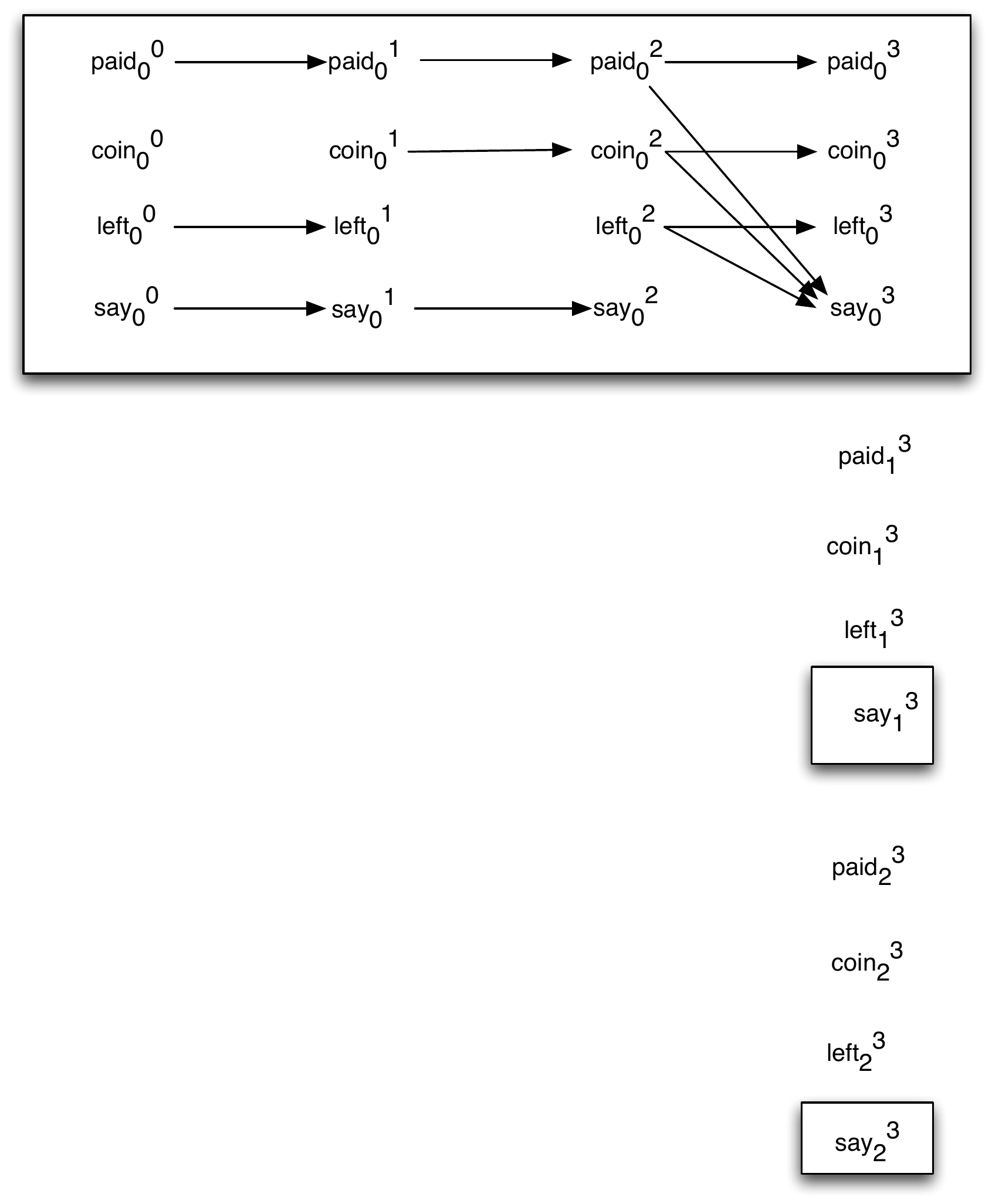}
}
\caption{Timed-variables used in algorithm of van der Meyden and Su. \label{fig:meydensu}} 
\end{figure}

From the point of model checking complexity, we expect that the simplification of the 
dependency graph will result in significant improved performance of the model checking computation. 
For $n$ cryptographers,  the initial dependency graph (Figure~\ref{fig:initgraph} for $n=3$) has $16 n$ variables, 
i.e., $48$ variables in case $n=3$. The algorithm of van der Meyden and Su \cite{MS} would construct a BDD with over
$12+4n$  variables in general, and, as show in Figure~\ref{fig:meydensu},  with 24 variables in case $n=3$. 
However, the algorithm uses an intermediate BDD representation of the transition relation of the 
protocol that requires $8n$ variables. Instead, the optimization 
approach developed here computes a BDD over just 9 variables in case $n=3$ and 
$3n$ variables in general. The actual model checking computation combines BDD's associated with each node
to construct a BDD over the same number of variables. 
Since in practice, BDD algorithms work for numbers of variables in the order of 100-200, 
these reductions of the constant factor can have a significant impact on the scale of the 
problems that can be solved. 

\section{Experimental Results} \label{sec:results} 

In the present section, we describe the results of a number of experiments designed to 
evaluate the performance of epistemic model checking using the conditional 
independence optimization, in comparison with the existing implementation in 
MCK. (Since MCK remains the only  symbolic epistemic model checker that deals with 
perfect recall knowledge, there are no other systems to compare to.) 
All experiments were conducted on  an Intel 2.8 GHz Intel Core i5 
processor with 8 GB 1600 MHz DDR3 memory running Mac OSX 10.10.

Except where indicated, the unoptimized model checking algorithm against which 
we compare is that invoked by the construct {\tt spec\_spr\_xn} in the MCK scripting language, 
which implements the algorithm of van der Meyden and Su \cite{MS}. (We refer to this algorithm as {\tt xn} in 
legends, and the algorithm using conditional independence optimization is referenced as {\tt ci}.) 
The results demonstrate both significant speedups of as large as four orders of magnitude, 
as well as a significant increase in the scale of problem that can be handled in 
a given amount of time.

{\bf Dining Cryptographers:} 
 Our first example is the Dining Cryptographers protocol \cite{chaum}, 
 discussed above.  It was first model checked using epistemic logic in \cite{MS}. 
This example scales by the number $n$ of agents; the number of state variables is $O(n)$, 
and the protocol runs for 3 steps. The initial condition needs to say that at most one
of the agents paid -- this is done by means of a formula of size $O(n^2)$. The rest of the script scales linearly. 
The formula in all instances  states that at time 3, agent C0 either knows that 
no agent pays, knows that C0 is the payer, or knows that 
one of the other agents is the payer, but does not know which. 
This involves $O(n)$ atomic propositions, and is of linear size in $n$.

\begin{figure} 
{\small 
\centerline{(a) \begin{tabular}[t]{|c || c | c |   c | c |   c | c |   c | c |   c | c |   c | c |   }  
\hline
$n$ &  3 & 4& 5& 6& 7& 8& 9& 10& 11& 12 \\ 
\hline 
xn & 0.1&  0.16&  0.37&  0.65&  1.34&  2.52&  37.92&  33.5&  846.19&  2775.4\\
\hline 
\end{tabular}}
}
\centerline{(b)~\includegraphics[height=6cm]{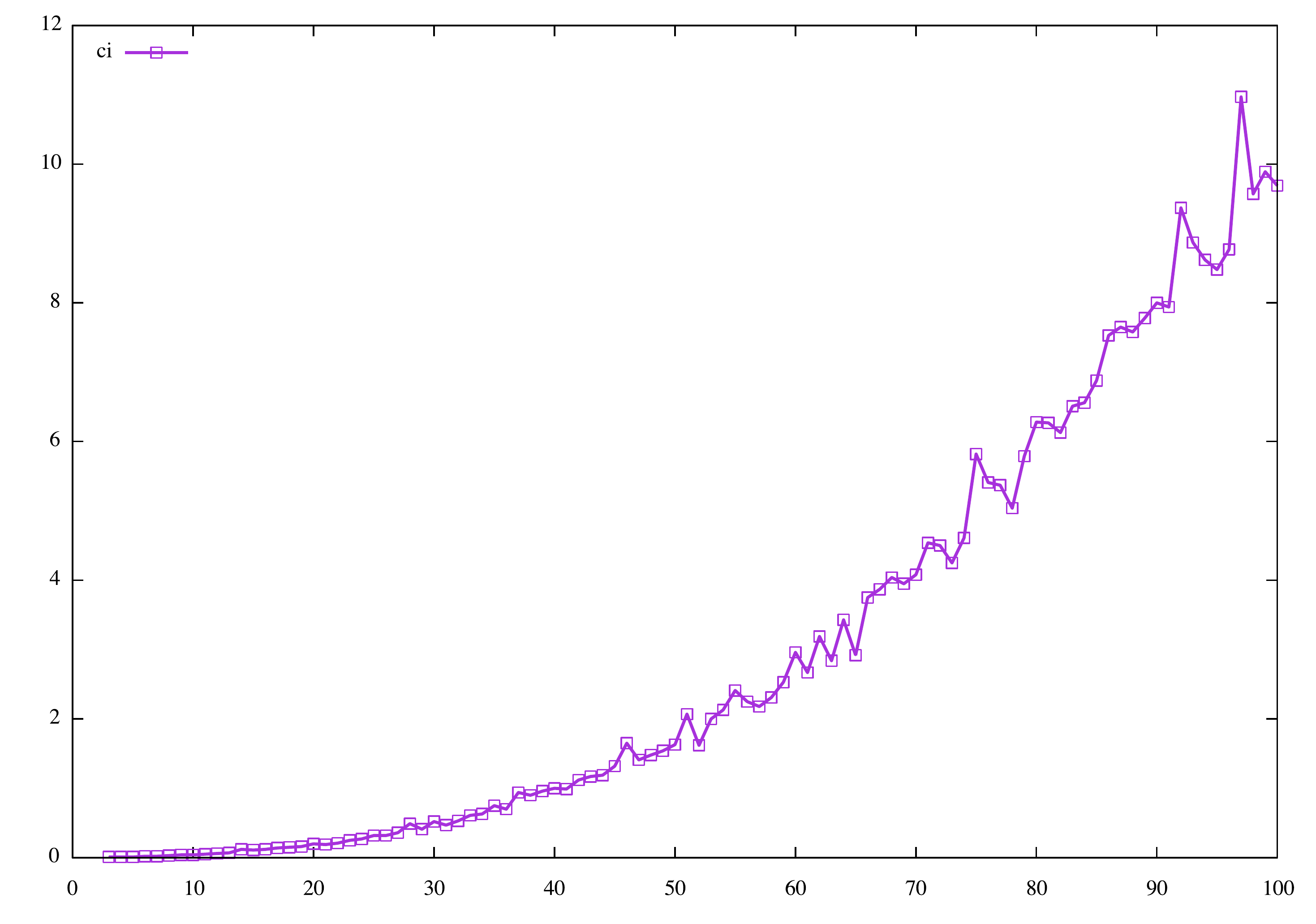}  }

\caption{\label{fig:dc:results} Dining Cryptographers experiments,
(a) unoptimized running times (s) and (b) 
 optimized  running times (s) } 
\end{figure}

Performance results for model checking the Dining Cryptographers protocol running on a ring with $n$ agents are
shown in Figure~\ref{fig:dc:results}(a). There is a rapid blowup as the number of agents is increased: 12 agents 
already takes over 46 minutes (2775 seconds). 
By contrast, applying the conditional independence optimization, 
model checking is significantly more efficient, as shown by the plot in 
Figure~\ref{fig:dc:results}(b). The case of 12 agents is handled in 0.05 seconds, 
and 100 agents are handled in 9.69 seconds.

{\bf One-time Pad:}  The next example concerns message transmission using one-time pad encryption in the presence of an eavesdropper.
Each instance has three agents (Alice, who sends an encrypted message to Bob,
and Eve, who taps the wire). We scale the example by the length of the message, which is sent one bit at a time.  
For a message of length $n$, states have $O(n)$ variables. 
The protocol runs $2n$ steps, two for each bit.  
The formula is evaluated at time $2n$, and 
says that Eve does not learn the value of the first bit. 
For this example, we found that the best performance for the unoptimized version was
obtained using MCK version 0.5.1, which used a different symbolic encoding of the transition relation 
from more recent versions.  Performance of model checking is shown in Table~\ref{fig:results-otp}. 

The running times for the optimized version grow very slowly.
Intuitively, the conditional independence optimization 
detects in this example that the first bit and the others are independent, and uses this 
to optimize the model checking computation. This means that for all $n$, the ultimate BDD model checking computation is 
performed on the same model for all $n$, and the primary running time cost lies in the generation of the 
dependence graph, and its analysis, that precedes the BDD computation. 
On the other hand, the unoptimized (xn) model checking running times show significant growth, with a large spike
towards the end, where the speedup obtained from the optimization is over 10,000 times. 

\begin{table} \centerline{
\begin{tabular}[b]{|c || c | c |   c | c |  }  
\hline 
$n$ & ci (s) & xn (s) & xn/ci \\
\hline  
11& 0.02& 1.25& 63\\
12& 0.02& 1.38& 69\\
13& 0.02& 2.04& 102\\
14& 0.02& 2.19& 110\\
15& 0.02& 3.98& 199\\
16& 0.03& 5.50& 183\\
17& 0.03& 5.01& 167\\
18& 0.03& 5.47& 182\\
19& 0.03& 7.24& 241\\
20& 0.04& 9.71& 243\\
\hline 
\end{tabular}
\begin{tabular}[b]{|c || c | c |   c | c |  }  
\hline 
$n$ & ci (s) & xn (s) & xn/ci \\
\hline  
21& 0.04& 8.42& 211\\
22& 0.04& 8.82& 221\\
23& 0.04& 11.10& 278\\
24& 0.04& 17.88& 447\\
25& 0.04& 36.68& 917\\
26& 0.04& 33.26& 832\\
27& 0.05& 23.60& 472\\
28& 0.05& 34.88& 698\\
29& 0.05& 99.50& 1990\\
30& 0.05& 50.10& 1002\\
\hline 
\end{tabular}
\begin{tabular}[b]{|c || c | c |   c | c |  }  
\hline 
$n$ & ci (s) & xn (s) & xn/ci \\
\hline  
31& 0.05& 75.13& 1503\\
32& 0.05& 67.37& 1347\\
33& 0.06& 97.23& 1621\\
34& 0.06& 184.19& 3070\\
35& 0.06& 89.47& 1491\\
36& 0.07& 131.74& 1882\\
37& 0.07& 164.76& 2354\\
38& 0.07& 259.48& 3707\\
39& 0.07& 275.87& 3941\\
40& 0.07& 749.88& 10713\\
\hline 
\end{tabular}
}
\caption{\label{fig:results-otp} One-time pad protocol, optimized  and unoptimized running times, and speedup ratio, ``single-bit'' formula} 
\end{table}

{\bf Oblivious Transfer:} 
 The next example concerns an oblivious transfer protocol due to Rivest \cite{rivest99}, 
which allows Bob to learn exactly one of Alice's two messages $m_0,m_1$, of his choice, without Alice knowing which
message was chosen by Bob.  
Each instance has two agents, and we scale by the length of the message. 
For a message of length $n$, states have $O(n)$ variables. 
We consider two formulas for this protocol. 
Both are evaluated at time 3 in all instances. 

The first formula says that if Bob chose to receive message $m_1$, then he does not learn the \emph{first} bit of $m_0$. 
The running times for model checking this formula are 
given in Table~\ref{tab:Rivest}(a). In this example, the 
conditional independence optimization gives a significant speedup, 
in the range of one to two orders of magnitude (more precisely, 12 to 221) 
improvement on the inputs considered, and increasing as the scale of the problem increases. 
Running just the optimized version on larger instances, we
find 
that the optimization allows us to handle significantly 
larger instances: up to 97 agents can be handled in under 200 seconds, 
compared with 19 agents in 170 seconds unoptimized. 

\begin{table*} \centerline{
\begin{tabular}[b]{|c || c | c |   c | c |  }  
\hline 
$n$ & ci (s) & xn (s) & xn/ci \\
\hline  
3& 0.02& 0.24& 12\\ 
4& 0.03& 0.52& 17\\
5& 0.05& 0.90& 18\\
6& 0.07& 1.80& 26\\
7& 0.11& 2.24& 20\\
8& 0.14& 3.54& 25\\
9& 0.15& 4.97& 33\\
10& 0.16& 7.20& 45\\
11& 0.21& 13.08& 62\\
12& 0.26& 16.68& 64\\
13& 0.32& 32.72& 102\\
14& 0.39& 62.08& 159\\
15& 0.43& 50.95& 118\\
16& 0.50& 36.73& 73\\
17& 0.60& 38.36& 64\\
18& 0.71& 69.27& 98\\
19& 0.77& 170.16& 221\\
20& 1.09& 148.56& 136\\
\hline 
\end{tabular}
~~~\begin{tabular}[b]{|c || c | c |   c | c |  }  
\hline 
$n$ & ci (s) & xn (s) & xn/ci \\
\hline 
3& 0.03& 0.25& 8.3\\
4& 0.05& 0.51& 10.2\\
5& 0.12& 0.86& 7.2\\ 
6& 0.15& 1.58& 10.5\\
7& 0.25& 2.84& 11.4\\
8& 0.42& 3.52& 8.4\\
9& 0.50& 5.11& 10.2\\
10& 0.55& 7.79& 14.2\\
11& 1.18& 13.07& 11.1\\
12& 3.72& 14.63& 3.9\\
13& 5.20& 39.74& 7.6\\
14& 7.13& 48.64& 6.8\\
15& 4.91& 56.62& 11.5\\
16& 20.16& 38.09& 1.9\\
17& 32.95& 42.40& 1.3\\
18& 174.96& 86.81& 0.5\\
19& 229.85& 96.86& 0.4\\
20& 342.40& 184.08& 0.5\\
\hline 
\end{tabular}
}
\centerline{(a)\hspace{4cm}(b)} 
\caption{\label{tab:Rivest} 
Rivest's Oblivious Transfer Protocol, 
(a)  ``single-bit'' formula, 
(b)  ``all bits'' formula
} 
\end{table*}

\begin{table*} \centerline{
\begin{tabular}[b]{|c || c | c |   c | c |  }  
\hline 
$n$ & ci (s) & nested(s) & nested/ci \\
\hline 
3& 0.01	& 0.02	& 2\\
4& 0.01	& 0.03	& 3\\
5 & 0.01	& 0.04 & 	4\\
6& 0.01	& 0.06& 	6\\
7& 0.01	& 0.11	& 11\\
8& 0.02	& 0.20	 & 10\\
9 & 0.03	& 0.46& 	15 \\
10 & 0.03	& 1.05	& 35 \\ 
11 & 0.05	& 2.44& 	49\\ 
12 & 0.07	& 5.69	& 81\\
13 & 0.09	& 14.5& 	161\\
14 & 0.12	& 34.77& 	290\\
15 & 0.16	& 89.31& 	558\\
16 & 0.20 & 	360.3& 	1802\\
17 & 0.27	& 1597.91	& 5918\\
\hline 
\end{tabular}
~~~
\begin{tabular}[b]{|c || c | c |   c |  }  
\hline 
$n$ & ci (s) & xn (s)  \\
\hline 
3  &  0.04 & 0.79\\ 
4  &  0.11 & 96.47\\ 
5 &  0.49 & $>$ 2 hrs \\ 
6 &  2.46 & - \\ 
7 & 12.93 & - \\ 
8  & 155.41 & - \\  
9 & $>$ 2hrs  & - \\
\hline 
\end{tabular}
}
\centerline{(a)\hspace{5cm}(b)} 
\caption{\label{tab:MTPC} 
(a)  Message Transmission Protocol, 
(b) Chaum's two-phase protocol.
} 
\end{table*} 

An example in which the optimization does
not always yield a performance improvement 
arises when we change the formula model checked in this 
example to one that states that if Bob chose to receive $m_1$, then he does
not learn the value of \emph{any} bit of $m_0$. The running times are shown in Table~\ref{tab:Rivest}(b).  
Here, the optimization initially gives a speedup of roughly one order of magnitude, but 
on the three largest examples, the performance of the unoptimized algorithm is better
by a factor of two. The lower size of the initial speedup, compared to the first formula,  can be explained from the
fact that there are obviously fewer variables that are independent of the second formula, since the 
formula itself contains more variables. 
(The ``all bits'' formula contains $O(n)$ rather than just one variable explicitly, but recall that 
knowledge operators implicitly  introduce more variables, 
so the ``first bit'' formula implicitly has $O(n)$ variables.) 
It is not immediately clear exactly what accounts for the switchover. 

{\bf Message Transmission:} 
The next example concerns the transmission of a single bit message 
across a channel that is guaranteed to deliver it, but with uncertain delay.
This example has two agents Alice and Bob , and runs for $n+1$ steps, where $n$ is the 
maximum delay.  States have $O(n)$ variables. The formula considered  
asserts at time $n+1$ that Alice knows that Bob knows ... (nested five levels) that the message  has arrived. 
Because of the nesting, the algorithm used in the unoptimized case is that 
invoked by the MCK construct {\tt spec\_spr\_nested} -- this essentially performs
BDD-based model checking in a structure in which the worlds are runs of length 
equal to the maximum time relevant to the formula. 

Table~\ref{tab:MTPC}(a) compares the performance of the conditional independence
optimization with this algorithm. The degree of optimization obtained is significant, 
increasing to over four orders of magnitude. 
Running just the optimization for larger instances, we find that 
the optimization enables significantly larger instances to 
be handled in a given amount of time: as many as 65 agents in 
342 seconds, compared to just 16 agents in 360 seconds for the unoptimized version.

{\bf Chaum's two-phase protocol:} 
The final example we consider is Chaum's two-phase protocol \cite{chaum},  a protocol for anonymous broadcast that 
uses multiple rounds of  the Dining Cryptographers protocol. 
Model checking of this protocol has previously been addressed 
in \cite{BatainehMeyden11}. 
This example scales by both the number of agents and the number of steps of the protocol: 
with $n$ agents, the protocol runs for $O(n)$ steps, and each state is comprised of 
$O(n)$ variables. We check a formula with $O(n)$ variables that says that the first agent 
has a bit {\tt rcvd1} set to true at the end of the protocol iff it knows that 
some other agent is trying to send bit 1. 
The protocol is more complex than the others considered above. 
An initial set of $n$  ``booking'' rounds of the Dining Cryptographers protocol is 
used to anonymously attempt to book one of $n$ slots, 
and this is followed by $n$ ``slot" rounds of the Dining Cryptographers protocol,  
in which an agent who has booked a slot without detecting a collision with another agent's booking, uses that slot to 
attempt to broadcast a message. Because undetected booking collisions remain possible, 
collisions might also be detected in the second phase. 
Because of the complexity of the protocol, this example can 
only be model checked on small instances in reasonable time, even 
with the optimization. 
Table~\ref{tab:MTPC}(b) shows the running times obtained: 
for the unoptimized version, we again used MCK-0.5.1.  
The running time of the unoptimized computation explodes at $n=5$ as we increase the 
number of agents. The optimized computation takes significantly less time, but also eventually explodes, at $n=9$.
Thus, the optimization has doubled the size of the problem that can be handled in reasonable time. \\[-20pt]

\section{Related Work and Conclusion} \label{sec:concl} 

We conclude with a discussion of some related work and future directions. 

Wilson and Mengin \cite{WM01} have previously related modal logic to valuation algebra, 
but their definition requires that the marginalization of a Kripke structure have exactly the 
same set of worlds and equivalence relation, and merely restricts the assignment at 
each world, so their approach does not give the optimization that we have developed, 
and a model checking approach based on it would be less efficient than that developed in the 
present paper. They do not discuss conditional independence, which is a key part of our approach. 

Also related are probabilistic programs, a type of program containing probabilistic choice statements, that
sample from a specified distribution. The semantics of such programs is that they generate a 
probability distribution over the outputs. 
These programs may contain statements of the form $observe(\phi)$ where $\phi$ is a boolean condition: 
these are interpreted as conditioning the distribution constructed to that point on the condition $\phi$. 
Hur et al.~\cite{HurNRS14} develop an approach to slicing probabilistic programs based on a static analysis that
incorporates ideas from the Bayesian net literature. 
There are several  differences between probabilistic programs and our work in this paper. 
One is that we deal with discrete knowledge rather than probability -- in general, 
this makes our model checking problem more tractable.  We also
reason about all possible sequences of observations, rather 
than one particular sequence of observations. 
Additionally, we allow observations by multiple agents rather than just one. 
Finally, via knowledge operators, we have a locus of reference to observations in our framework that is located in 
formulas rather than inside the program -- this enables us to ask multiple questions about a program 
without changing the code, whereas in probabilistic programs, one would need to handle this
by multiple distinct modifications of the code. 

The results of the present paper concern formulas that refer (directly and through knowledge operators) only
to a specific time. Our approach, however, can be easily extended by means of a straightforward  
transformation to formulas that talk about multiple time points, and we intend to implement
this extension in future work. 
The technique we have developed can also be extended to deal with multi-agent models 
based on programs taking probabilistic transitions, which MCK already supports. 
Formulas in this extension would include operators that talk about an 
agent's subjective probability, given what it has observed. 

Other extensions we intend to implement are to enrich the range of knowledge semantics
beyond the synchronous perfect recall semantics treated in this paper: essentially 
the same techniques will apply to the clock semantics (in which an agent's knowledge is based on 
just its current observation and the current time). The observational semantics, in which the agent's knowledge
is based just on its current observation, will be more challenging, since it is asynchronous, and 
knowledge formulas may refer to times arbitrarily far into the future. 
 
\bibliographystyle{eptcs} 
\bibliography{condind}
\end{document}